\documentclass[11pt,twoside]{article}
\usepackage{graphicx}

\usepackage{asp2006}

\begin{document}
\title{A journey across the M33 disk}   
\author{Edvige Corbelli, Laura Magrini, Simon Verley}   
\affil{INAF-Osservatorio Astrofisico di Arcetri, Firenze, ITALY}   

\begin{abstract}
The Local Group member M33 is a pure disk galaxy bearing
no prominent bulge or stellar halo. It constitutes a challenge for any 
hierarchical galaxy formation theory and an ideal laboratory for studying
quiescent star formation. 
Using multiwavelength observations of the gas and stellar component
in this nearby galaxy we are able to constrain the gas accretion and
star formation history. In the centermost region we find
kinematical evidence of a weak bar, which explains the central light 
excess and the enhanced metallicity. In the more extended disk the lack
of strong gradients of metal and dust abundances supports the picture that
the slow radial decline of the star formation rate is due to a change in the 
large scale disk perturbations: bright HII regions and giant molecular clouds
being born only in the inner disk. 
The analysis of the infrared Spitzer maps has however revealed hundreds of low 
luminosity star forming sites in places with a variety of dust content.
These are essential ingredients for understanding the overall gas to star 
formation process in M33 and in more distant late type galaxies.

\end{abstract}


\section{Introduction}  
Our knowledge of  the processes in the interstellar medium
that favor the birth of stars is mostly based on Galactic studies and 
on luminous, gas rich galaxies with high star formation rates.   
Scaling factors across our galactic disk are difficult to
estimate, being our Sun deeply embedded in it.
Imaging the gas and the newly star forming sites in external galaxies 
with high sensitivity and resolution is necessary in order to evaluate 
whether the ingredients necessary to trigger star formation 
vary along the Hubble sequence, to identify 
the role of large scale perturbations and to define the   
relationship of newly born stars to other ISM components across the disk.
Before the next generation of telescopes will resolve distant galaxies 
with a wide range of physical and dynamical properties, it is desirable 
to focus on the closest galaxies. M33, at a distance of 840~kpc
(Freedman, Wilson $\&$ Madore 1991), has a high star formation rate per 
unit area compared to M31 and a low
extinction towards star forming regions. It bears no prominent 
bulge and no signs of recent mergers. In addition, star counts in the outer 
disk indicate that any stellar halo component contributes for less than a few
percent to the total luminosity (Barker et al. 2006; McConnachie et al. 2006). 
It is therefore a prototype for any meeting focused on `disk galaxies', 
a challenge for any hierarchical galaxy formation theory, 
and a reference point for any evolutionary scenarios involving blue,  
low luminosity objects. Being relatively undisturbed,
its blue color and prominent HII regions require some quiescent feeding 
mechanisms to sustain the extended star formation activity.

We outline here some aspect of the gas pathway to stars:
from the needs of gas accretion to the formation of molecular clouds,
from the imprints of a weak central bar to properties of the outermost star 
forming sites. 
 
\section*{Fueling star formation across the disk}

The innermost 1~kpc region of M33 deserves particular attention of the
because the enhancement in the light distribution, metal abundance and 
molecular hydrogen surface density. Recent optical observations
of gas and stellar radial velocities in this region (Corbelli $\&$ Walterbos
2007) show ordered but non-circular motion down to about 1$''$ from the center. 
The most likely explanation for the observed velocity patterns are streaming 
motions along a weak inner bar with a position angle close to that of the minor 
axis, consistent with the central enhancements.

Using the rotational velocity gradient and the surface density of baryonic
matter, derived by the mass model fit to rotational data, it is possible 
to explain the 7.5~kpc size of the star forming disk of M33. The disk is in fact
gravitationally unstable for $R<7.5$~kpc according to the Toomre stability
criterion for a two component fluid: the gas and the stars (Corbelli 2003).
Inside this region and excluding the innermost 1~kpc, the 
radial variations of several quantities are very shallow: the atomic 
to molecular gas ratio is $\propto R^{0.6}$  (Heyer et al. 2003), the 
O/H metallicity gradient is $\le 0.07$~dex kpc$^{-1}$ (Magrini et al. 2007; 
Rosolowsky $\&$ Simon 2007), the gas surface density and the dust to gas ratio 
are almost constant (Verley et al. 2008, to be submitted). However the spiral
pattern is much more prominent for $R<4$~kpc and most of the
giant molecular clouds lie inside this region. One possibility is that
the fading of the spiral pattern, coincident with the corotation radius or with 
the outer Lindbland resonance of the bar, limits the formation of giant clouds
and large stellar complexes. 
A comparison between single dish and interferometric
surveys of the CO J=1-0 line moreover shows that Giant Molecular clouds 
contain only a small fraction of the total molecular mass 
(Corbelli 2003, Engargiola et al. 2003, Heyer et al. 2004, Gardan et
al. 2007) and that the mass spectrum is much steeper than our Milky Way 
(Blitz $\&$ Rosolowsky 2005).  
The atomic to molecular gas conversion in a galaxy 
with a low molecular fraction like M33 (about $10\%$ of the total 
gas mass inside the star forming disk is molecular) is   
regulated by the balance between the interstellar radiation field and
the hydrostatic pressure due to stars and gas (Elmegreen 1993; Heyer et
al. 2003). Fragmentation then locks most of the mass in small clouds.
It is still unclear whether these are all gravitationally bound units
but they are the main triggers of star formation at large radii.

A chemical evolution model for M33, constrained by the distribution of
stars, gas, and chemical abundances in stellar populations of different
ages, shows that a continuous gas infall of about 1~M$_\odot$~yr$^{-1}$ is
needed to fuel a star formation and that this is slowly declining with 
cosmic time (Magrini, Corbelli, $\&$ Galli 2007). 
The model and 21-cm observations (Westmeier, Braun $\&$ Thilker 2005) 
favor an extended phase for the inside-out formation of the disk, with 
the intergalactic medium providing the gas supply to it. 

\section*{The Cluster Birthline}

From the 24$\mu$m Spitzer map of M33 (Verley et al. 2007) we extracted
hundreds of sources which have H$\alpha$ counterpart
(Hoopes $\&$ Walterbos 2000). Most of these sources are likely to be
young stellar clusters, and the proximity of M33 gives the possibility
of selecting even the low luminosity ones, which contain only single OB 
associations, or clusters not massive enough to have made an O-type star.
We can look in detail to the properties of sites where 
clusters of different masses are born and evolve. 
Verley et al (2007) have already shown that the relation between the 
infrared luminosity and H$\alpha$ for these sources has a large scatter.
In general, if one looks to young clusters of different masses the
expected relation is far from
being linear and depends on the Initial Mass Function and on the fraction
of the bolometric luminosity absorbed by dust and re-emitted in the infrared.
Figure 1 shows this expected relation for a Salpeter IMF (dashed line) 
under the assumption that most of the cluster bolometric luminosity is
radiated by the surrounding dust in the infrared. We shall call this
relation {\it The Cluster Birthline}. From left to right the
cluster luminosity as well as its mass and its maximum stellar
mass increase.  Star symbols indicates the corresponding values for single
main sequence stars, of spectral type equal or later than O3.
We infer the total infrared luminosity (between 3-1100 $\mu$m) 
using Eq. (1) of Calzetti et al. (2006)  and an $8\mu$m/24$\mu$m flux 
ratio equal to unity, as it is on average in M33 
(i.e. $L(TIR)={\hbox{log}} L(24) + 0.908$).
Data should all lie  above the sequence 
marked by the star symbols and mostly around the Cluster Birthline if the
dominant population are young star clusters. Any aging, extinction, as well as 
the loss of ionizing photons leaking out from the HII region will bring the
data points above it. Instead there are few points above the birthline
and many below it. The conclusion is that the abundance of dust around those 
sources is low and varying and that the bolometric luminosity of most sources 
is recovered only complementing the IR photometry with UV and optical data.
This will be shown in detail in a forthcoming paper.
The fraction of bolometric luminosity absorbed by grains and re-emitted at 
IR wavelengths is higher for the most luminous sources which are surrounded 
by high column density gas 
(filled points in the figure with $N_{Htot}>10^{21}$~cm$^{-2}$).
The absence of sources in the bottom right corner of the figure is
due to the absence of stars more massive than O3 type stars in our 
sample.

\section*{Conclusions}

The M33 disk is made up of four distinct zones: the
central one where there is kinematic evidence for 
a weak bar. This limit the ability to constrain the dark 
matter density profile in its center and in more distant, less resolved
galaxies. The second zone is the inner disk, where spiral arms trigger 
the formation of giant molecular clouds and of bright star 
forming sites. The third one, at larger galactocentric radii,  
where active star formation  proceeds in smaller subunits. These
give a substantial contribution to the global star formation rate and
have a patchy dust to gas ratio. And finally the outer regions. Here 
a warped outer disk is in place, linking the active 
star forming disk to the environment, to the surrounding intergalactic
space, which is providing the fuel to star formation.

\acknowledgements 
We would like to acknowledge D. Galli, C. Giovanardi, M. Grossi, L. Hunt,   
and R. Walterbos as collaborators and members of the FM33 (Friends of M33) 
working group and P. Lenzuni for evening discussions on the Cluster Birthline.

\begin{figure}
\vspace{7.0cm}
\includegraphics{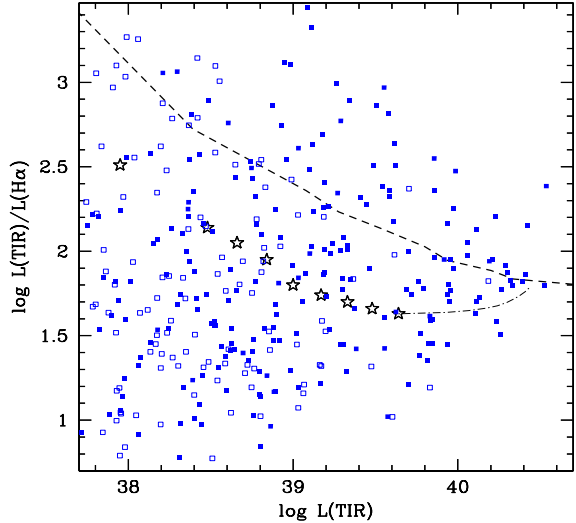}
\caption{Total Infrared Luminosity (TIR) of 24$\mu$n selected sources with 
H$\alpha$ counterpart (square symbols)  versus the TIR/H$\alpha$ luminosity
ratio. The dashed curve is the Cluster Birthline, where young stellar clusters
with a Salpeter IMF should lie. Filled square symbols are for sources in a 
high column density environment. Open star symbols are the expected values for single 
main sequence stars. The dash-dotted line connects the expected values for 
single O3 stars to clusters which have their maximum stellar mass equal to
an O3-type star. See Section 3 for more details. }
\label{fig1}
\end{figure}


\begin{thebibliography}{}
\bibitem[]{bar06}Barker, M. K., Sarajedini, A., Geisler, D., Harding, P.,
       \& Schommer, R. 2006, Astron. J., 133, 1138
\bibitem[]{bli05}Blitz, L., $\&$ Rosolowsky, E. 2005 in The Initial Mass Function 
      50 years later. eds. E. Corbelli, F. Palla, \& H. Zinnecker, 
      ASSL 327 (Springer:Dordrecht), p.287
\bibitem[]{cal05}Calzetti, D. et al. 2005, ApJ, 633, 871
\bibitem[]{cor07}Corbelli, E., \& Walterbos, R.A.M. 2007, ApJ, 669, 315
\bibitem[]{elm93}Elmegreen, B. G. 1993, ApJ, 411, 170
\bibitem[]{eng03}Engargiola, G., Plambeck, R. L., Rosolowsky, E. $\&$ Blitz, L.
      2003, ApJS, 149, 343
\bibitem[]{fre91}Freedman, W. L., Wilson, C. D., $\&$ Madore, B. F. 1991, 
         ApJ, 372, 455
\bibitem[]{gar07} Gardan, E., Braine, J., Schuster, K. F., Brouillet, N., 
         \& Sievers, A., 2007, A$\&$A, 473, 91
\bibitem[]{hey04}Heyer, M. H., Corbelli, E., Schneider, S. E., \& Young, J. S.
             2004, ApJ, 602, 723
\bibitem[]{hoo00}Hoopes, C. G., $\&$ Walterbos, R. A. M. 2000, ApJ, 541, 597
\bibitem[]{mag07}Magrini, L., Corbelli, E., \& Galli, D. 2007, A\&A, 470, 843
\bibitem[]{mcc06}McConnachie, A. W., Chapman, S. C., Ibata, R. A., et al.
          2006, ApJ, 647, 25
\bibitem[]{ros07}Rosolowsky, E., $\&$ Simon, J. D., 2007, ApJ, in press
\bibitem[]{ver07}Verley, S., Hunt, L. K., Corbelli, E., \& Giovanardi, C. 2007,
             A\&A, 476, 1161
\bibitem[]{wes05}Westmeier, T., Braun, R., Thilker, D. 2005, A\&A, 436, 101
\end{thebibliography}
\end{document}